\documentclass[final,3p,times]{elsarticle}

\usepackage{amssymb}
\usepackage{amsmath}
\usepackage{graphicx}
\usepackage{hyperref}
\usepackage{url}

\journal{Journal of Subatomic Particles and Cosmology}

\begin{document}

\begin{frontmatter}

\title{News on charged and neutral hadron production from NA61/SHINE}

\author[ncbj]{Yuliia Balkova}
\author[ifj]{Maciej Lewicki\corref{cor1}}
\ead{maciej.lewicki@ifj.edu.pl}
\cortext[cor1]{Corresponding author. Presenting author.}
\author{for the NA61/SHINE Collaboration}

\affiliation[ncbj]{organization={National Centre for Nuclear Research},
             city={Warsaw},
             postcode={02-093},
             country={Poland}}

\affiliation[ifj]{organization={Henryk Niewodnicza{\'n}ski Institute of Nuclear Physics, Polish Academy of Sciences},
             city={Krakow},
             postcode={31-342},
             country={Poland}}

~\\[-15mm]

\begin{abstract}
Strangeness production in high-energy hadronic and nuclear collisions remains a central topic in the study of strongly interacting matter under extreme conditions.
The data collected by the NA61/SHINE experiment at the CERN SPS allow for a comprehensive scan of hadron production across various collision energies and system sizes.

This article focuses on new results on charged and neutral hadron production in central collisions of medium-sized nuclei, such as Ar+Sc and Xe+La, in the SPS energy range.
First, the energy and system-size dependence of the $K^+/\pi^+$ ratio is discussed, confirming the ``horn'' structure in central Pb+Pb collisions and its absence in smaller systems.
Furthermore, the inverse slope parameter $T$ of kaon transverse mass (transverse momentum) spectra is analyzed, showing that intermediate-size systems reach values comparable to those observed in central Pb+Pb collisions.
Second, normalized proton rapidity spectra $(\text{d}n/\text{d}y)/\langle W \rangle$ in central Be+Be, Ar+Sc, and Pb+Pb collisions are presented, demonstrating energy-dependent baryon transport as a function of system size.
Building on this baseline, $\Lambda$ hyperon production in central Ar+Sc collisions is explored across SPS energies, where rapidity spectra, mean multiplicities ($\langle \Lambda \rangle$), and ratios to pions are compared with model predictions and existing data.
Finally, an anomalous excess of charged over neutral kaons ($R_K > 1$) in central Ar+Sc collisions is reported, highlighting a potential unexpectedly large violation of isospin symmetry.
\end{abstract}


\end{frontmatter}

\vspace{-0.5cm}
\section{Introduction}
\label{sec:intro}
The study of strongly interacting matter under extreme conditions of high temperature and baryon density provides valuable insights into the transition between hadronic matter and the Quark-Gluon Plasma (QGP).
One of the earliest proposed signatures of QGP formation is the enhancement of strangeness production relative to baseline proton-proton and proton-nucleus collisions~\cite{Rafelski:1982pu}.
The NA49 experiment at the CERN Super Proton Synchrotron (SPS) observed a striking non-monotonic behavior in the energy dependence of the $K^+/\pi^+$ ratio in central Pb+Pb collisions, collectively referred to as the ``horn'' structure~\cite{Alt:2008zz}, which has been interpreted as a signature of the onset of deconfinement.

To systematically investigate the phase diagram of strongly interacting matter, the NA61/SHINE \cite{Abgrall:2014xwa} experiment has performed a scan in collision energy and system size.
A recent review of these measurements and findings is given in Ref.~\cite{Stefanek:2026ltn}.
The experiment employs a fixed-target setup with a large-acceptance hadron spectrometer, featuring high-resolution Time Projection Chambers (TPCs) and Time-of-Flight (ToF) detectors.
This configuration enables excellent tracking and momentum resolution, making a wide range of rapidity accessible, down to zero transverse momentum ($p_\text{T} = 0$).
Charged particles are identified using the measurement of specific energy loss ($\text{d}E/\text{d}x$) in the TPC gas and time-of-flight information, while neutral hadrons, such as $\Lambda$ hyperons and $K^0_S$ mesons, are reconstructed through their weak decay topologies.

\section{Strangeness production and system-size dependence of the ``horn''}
\label{sec:horn}
\begin{figure*}[ht]
\centering
\vspace{-0.2cm}
\begin{minipage}{0.27\textwidth}
  \centering
  \includegraphics[width=\textwidth]{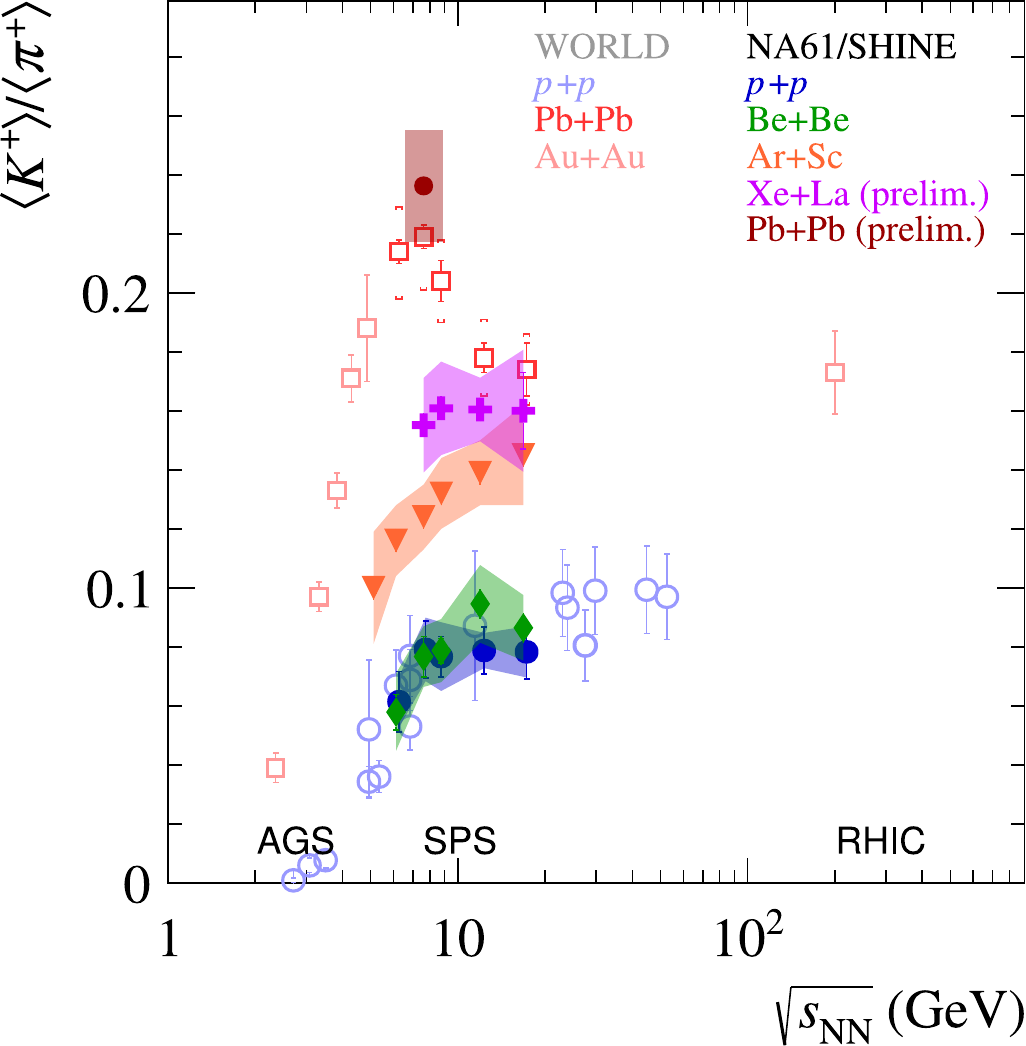}
\end{minipage}\hfill
\begin{minipage}{0.30\textwidth}
  \centering
  \includegraphics[width=\textwidth]{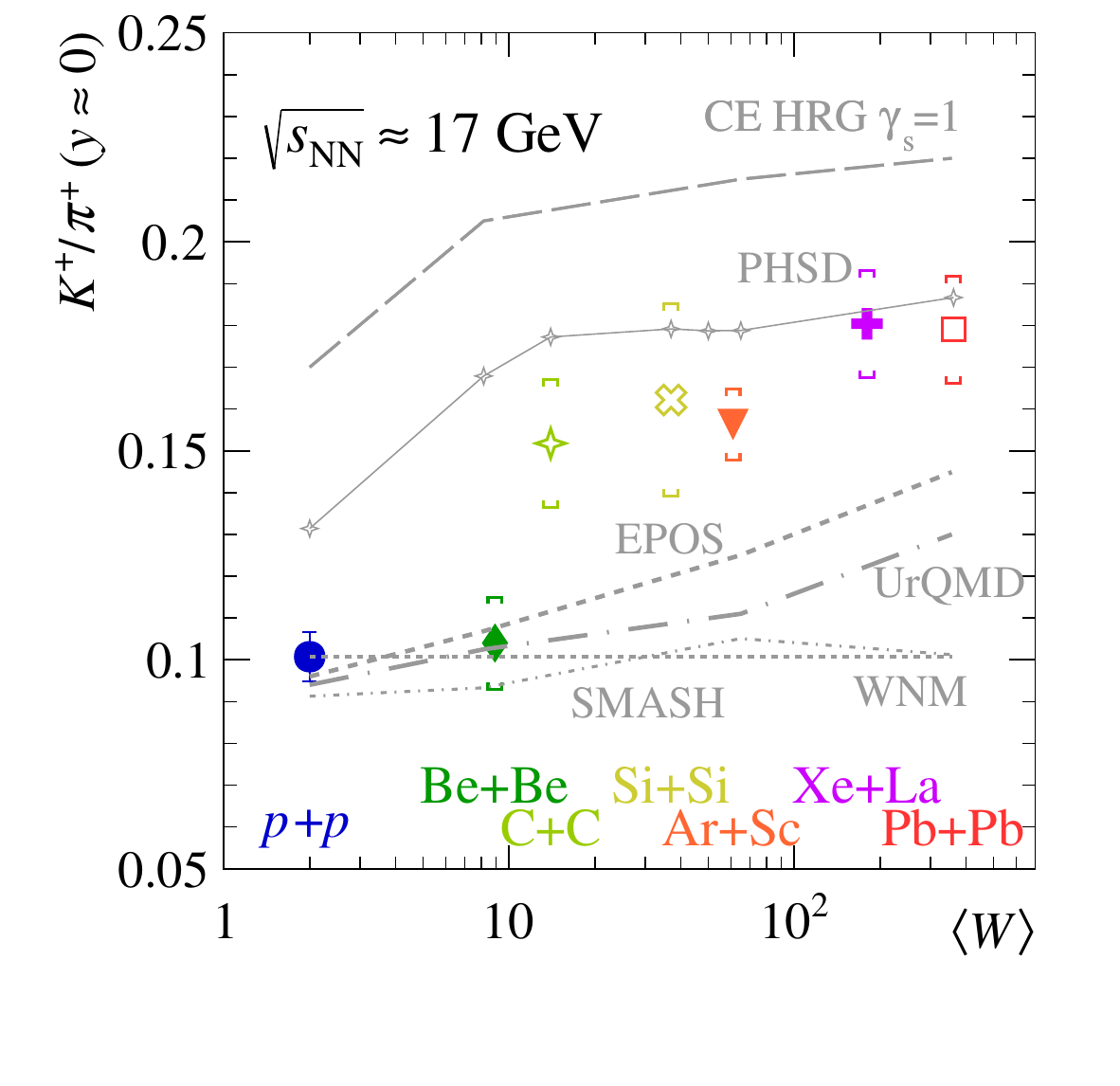}
\end{minipage}\hfill
\begin{minipage}{0.30\textwidth}
  \centering
  \includegraphics[width=\textwidth]{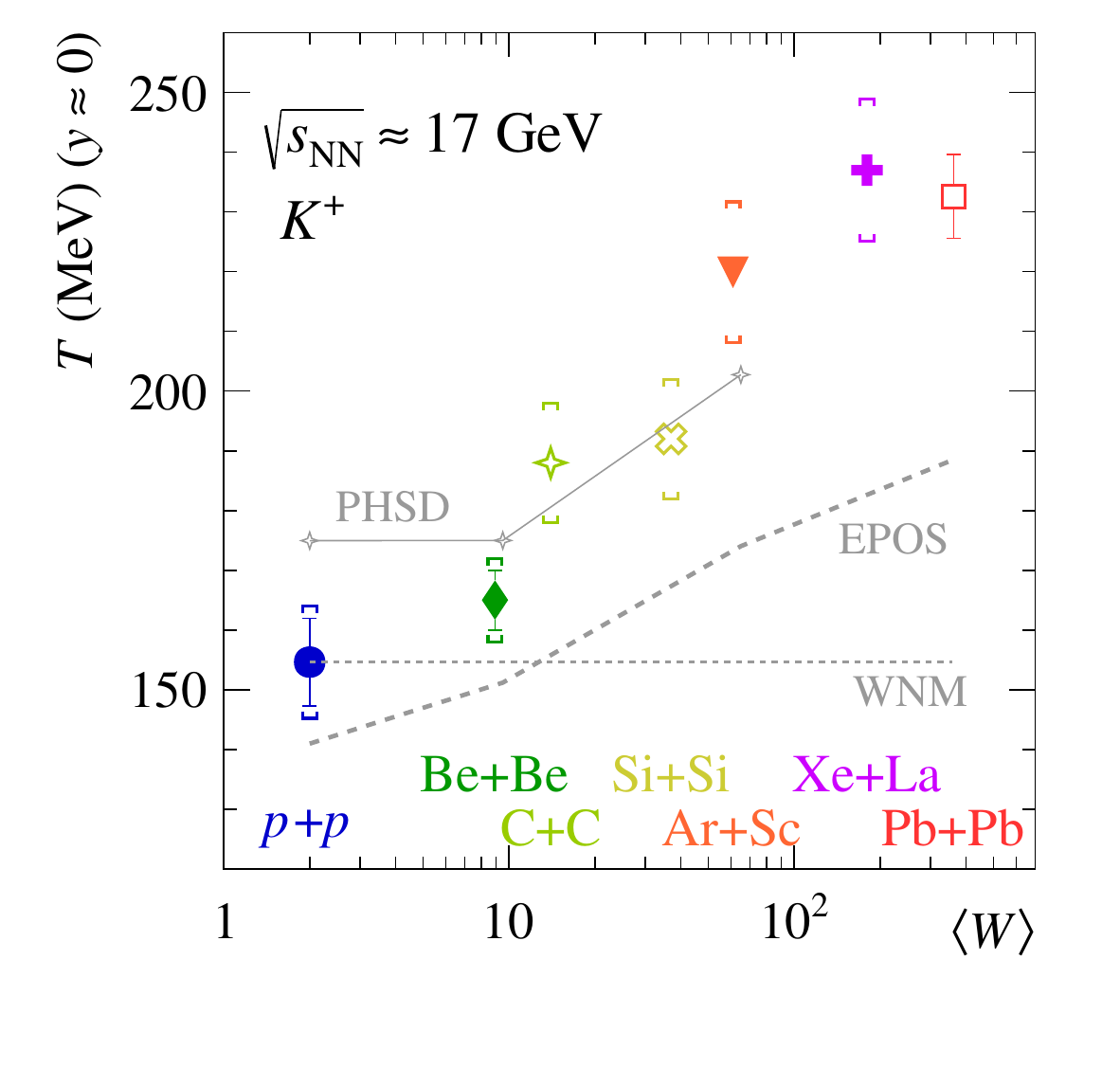}
\end{minipage}
\vspace{-0.4cm}
\caption{\textit{Left}: Energy dependence of the $\langle K^+\rangle/\langle\pi^+\rangle$ ratio in central Pb+Pb/Au+Au collisions compared with smaller systems (\textit{p+p}, Be+Be, Ar+Sc, Xe+La).
\textit{Middle} and \textit{right}: System-size dependence of the $K^+/\pi^+$ ratio (\textit{middle}) and the inverse slope parameter $T$ (\textit{right}) of $K^+$ at $y\approx0$ at $\sqrt{s_{\mathrm{NN}}}\approx17$~GeV in central Be+Be, C+C, Si+Si, Ar+Sc, Xe+La, and Pb+Pb collisions as well as inelastic \textit{p+p} interactions (NA61/SHINE: \textit{p+p}, Be+Be, Ar+Sc, Xe+La, Pb+Pb; NA49: C+C, Si+Si, Pb+Pb; STAR: Au+Au; AGS experiments: Au+Au; bubble chamber experiments: \textit{p+p}).
The system size is represented by the mean number of wounded nucleons $\langle W \rangle$.
For the left panel, statistical uncertainties for NA61/SHINE points are shown as error bars and systematic as shaded bands.
For the middle and right panels, statistical uncertainties are shown with bars and systematic uncertainties with square braces.
Grey lines represent model simulations.
For Xe+La, $y = 0.4 - 0.6$ was used as mid-rapidity.
See Ref.~\cite{NA61SHINE:2023epu} for references to published data and model predictions.}
\label{fig:horn_strangeness}
\end{figure*}

The ratio of $4\pi$ mean multiplicities $\langle K^+\rangle/\langle\pi^+\rangle$ is a sensitive proxy for the strangeness-to-entropy ratio in the system \cite{NA35:1994veo, Gazdzicki:1996pk}.
The left panel of Figure~\ref{fig:horn_strangeness} shows the energy dependence of the $\langle K^+\rangle/\langle\pi^+\rangle$ ratio in central Pb+Pb/Au+Au collisions compared with smaller systems.
The characteristic sharp maximum in the $K^+/\pi^+$ ratio (the ``horn'') was originally observed by NA49~\cite{Alt:2008zz} and supported by STAR results~\cite{STAR:2017epu} in central heavy-ion collisions.
This non-monotonic energy dependence is now confirmed by the preliminary measurement by the NA61/SHINE Collaboration in central Pb+Pb collisions at 30$A$ GeV/$c$.
In contrast, lighter systems (\textit{p+p}, Be+Be, Ar+Sc, Xe+La) exhibit no such peak, and a monotonic increase of the ratio is observed, with values for Ar+Sc and Xe+La lying between those of \textit{p+p}/Be+Be and Pb+Pb.

The transition from small to large systems is investigated using the $K^+/\pi^+$ ratio and the inverse slope parameter $T$ of $K^\pm$ transverse mass (transverse momentum) spectra at mid-rapidity ($y \approx 0$).
Figure~\ref{fig:horn_strangeness} shows these quantities as a function of the mean number of wounded nucleons $\langle W \rangle$ at $\sqrt{s_{\mathrm{NN}}}\approx17$~GeV (middle and right panels, respectively).
A rapid increase of both quantities is observed when moving from \textit{p+p} to Ar+Sc, followed by a slower increase or saturation towards Pb+Pb.
None of the applied theoretical models can reproduce the data across the entire range of system sizes, indicating that particle production and system-size evolution are not fully described by existing model frameworks.

\vspace{-0.6cm}
\section{System-size dependence of proton rapidity spectra and baryon transport}
\label{sec:protons}
\begin{figure*}[ht]
\centering
\begin{minipage}{0.25\textwidth}
  \includegraphics[width=\textwidth, trim=0cm 0cm 0cm 1.5cm, clip]{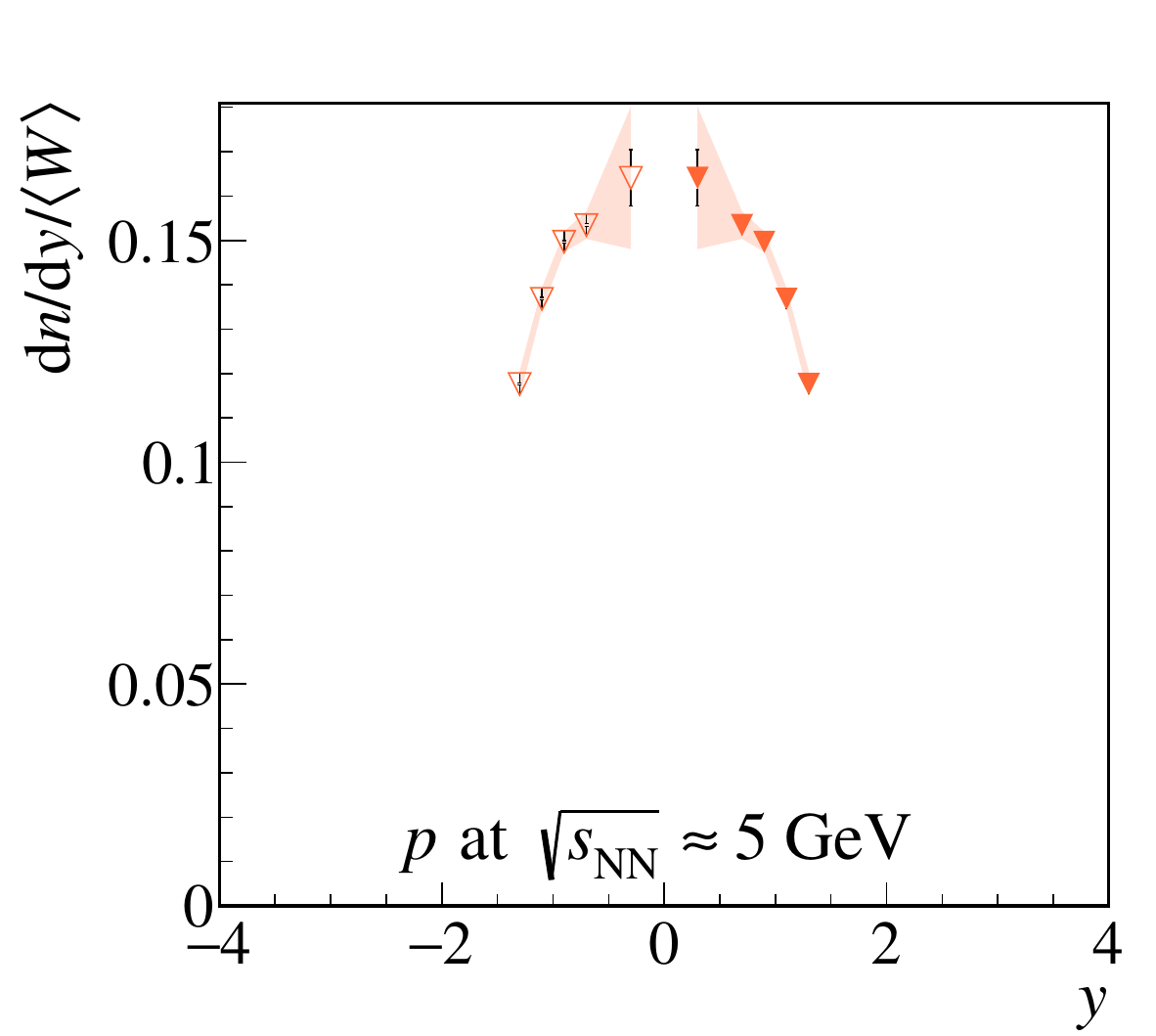}
\end{minipage}
\begin{minipage}{0.25\textwidth}
  \includegraphics[width=\textwidth, trim=0cm 0cm 0cm 1.5cm, clip]{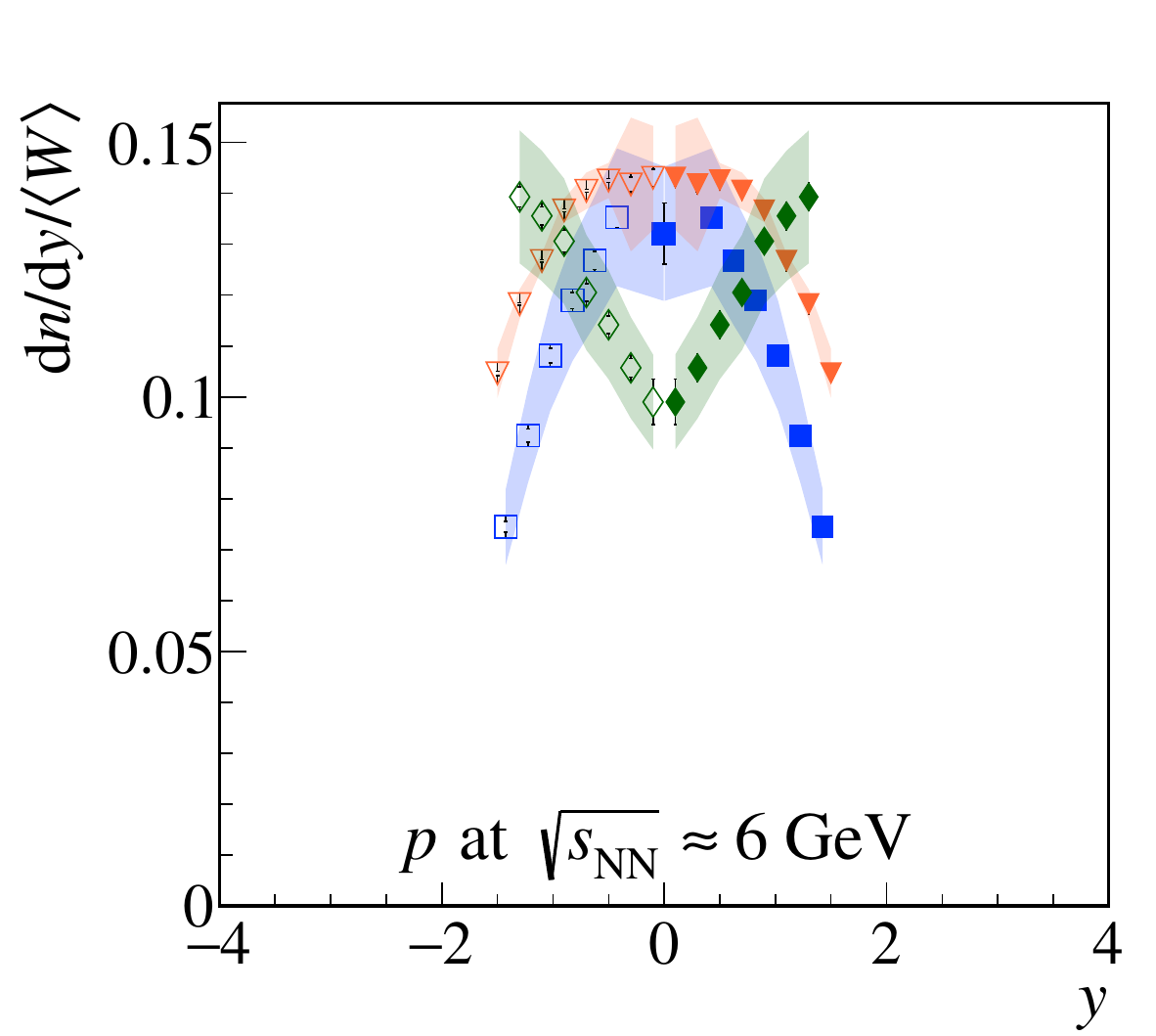}
\end{minipage}
\begin{minipage}{0.25\textwidth}
  \includegraphics[width=\textwidth, trim=0cm 0cm 0cm 1.5cm, clip]{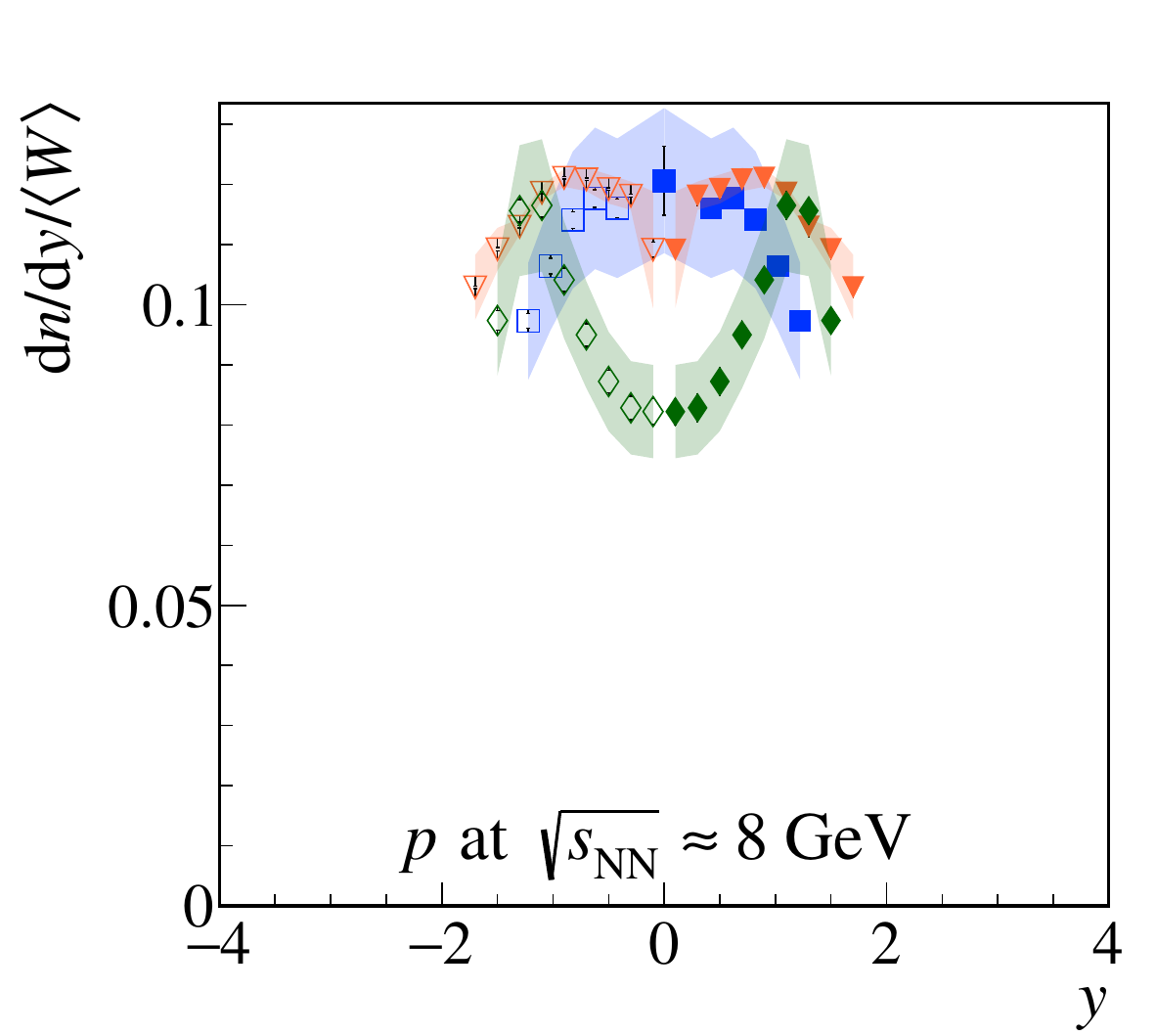}
\end{minipage} 
\begin{minipage}{0.25\textwidth}
  \includegraphics[width=\textwidth, trim=0cm 0cm 0cm 1.5cm, clip]{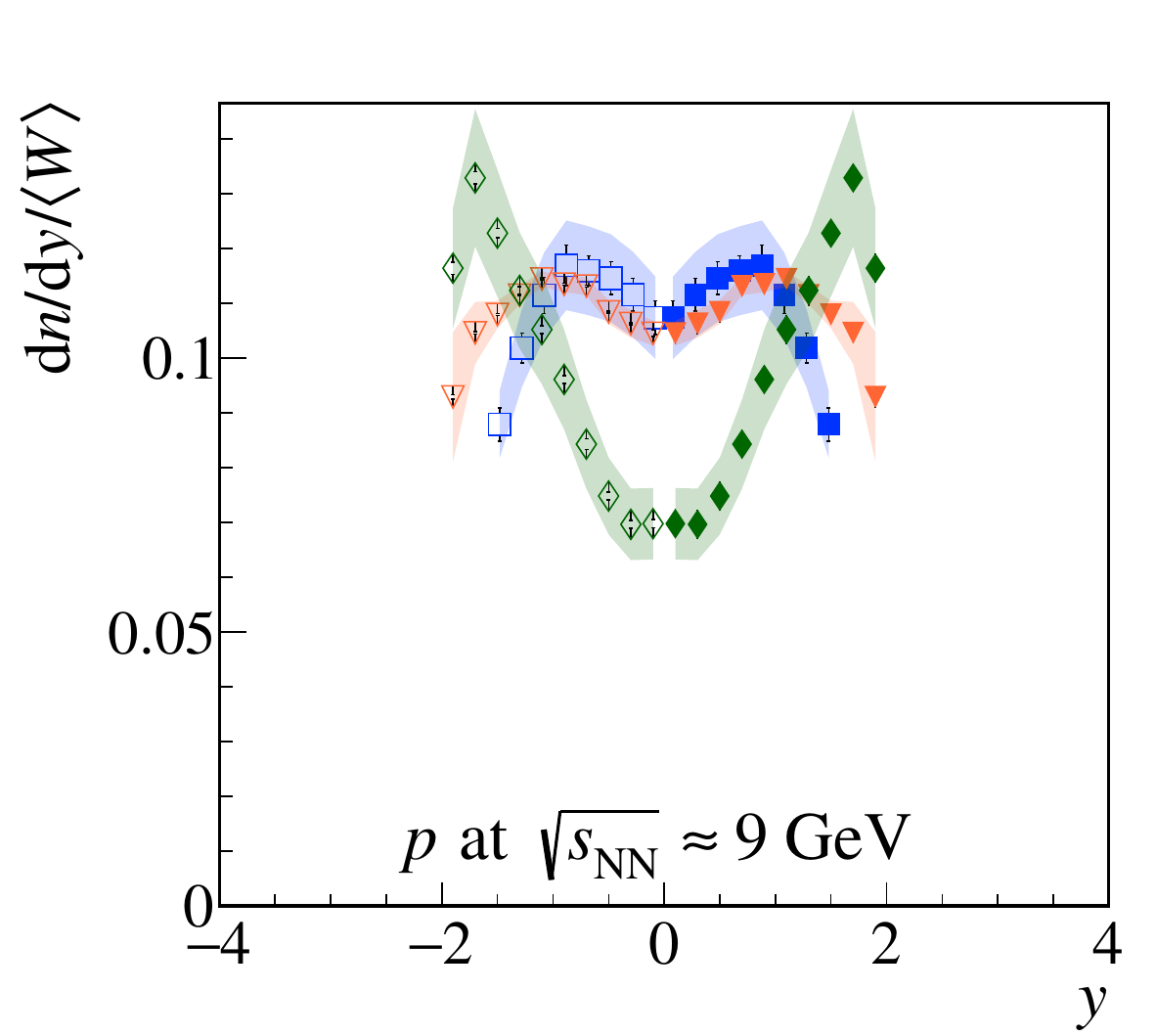}
\end{minipage}
\begin{minipage}{0.25\textwidth}
  \includegraphics[width=\textwidth, trim=0cm 0cm 0cm 1.5cm, clip]{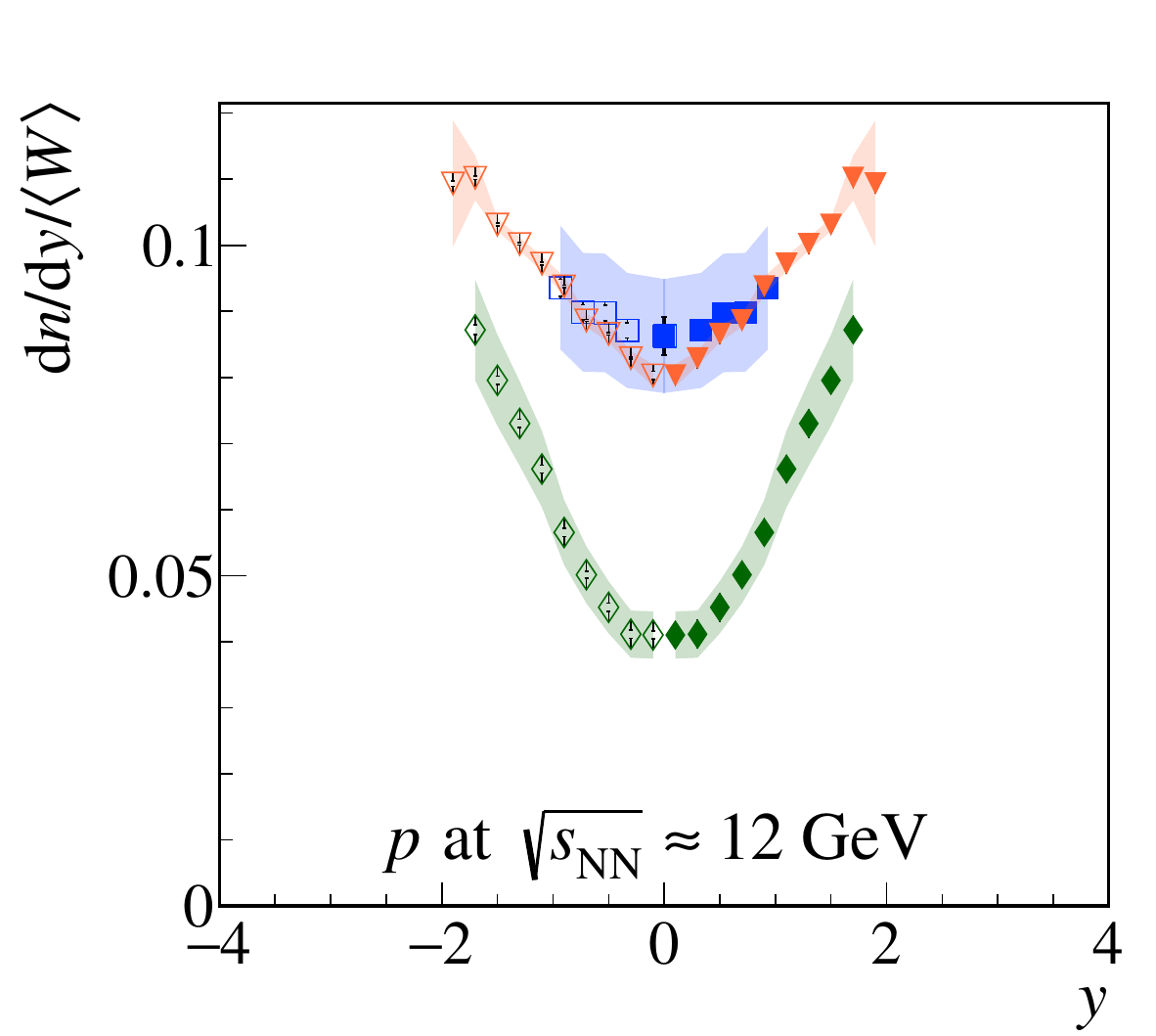}
\end{minipage}
\begin{minipage}{0.25\textwidth}
  \includegraphics[width=\textwidth, trim=0cm 0cm 0cm 1.5cm, clip]{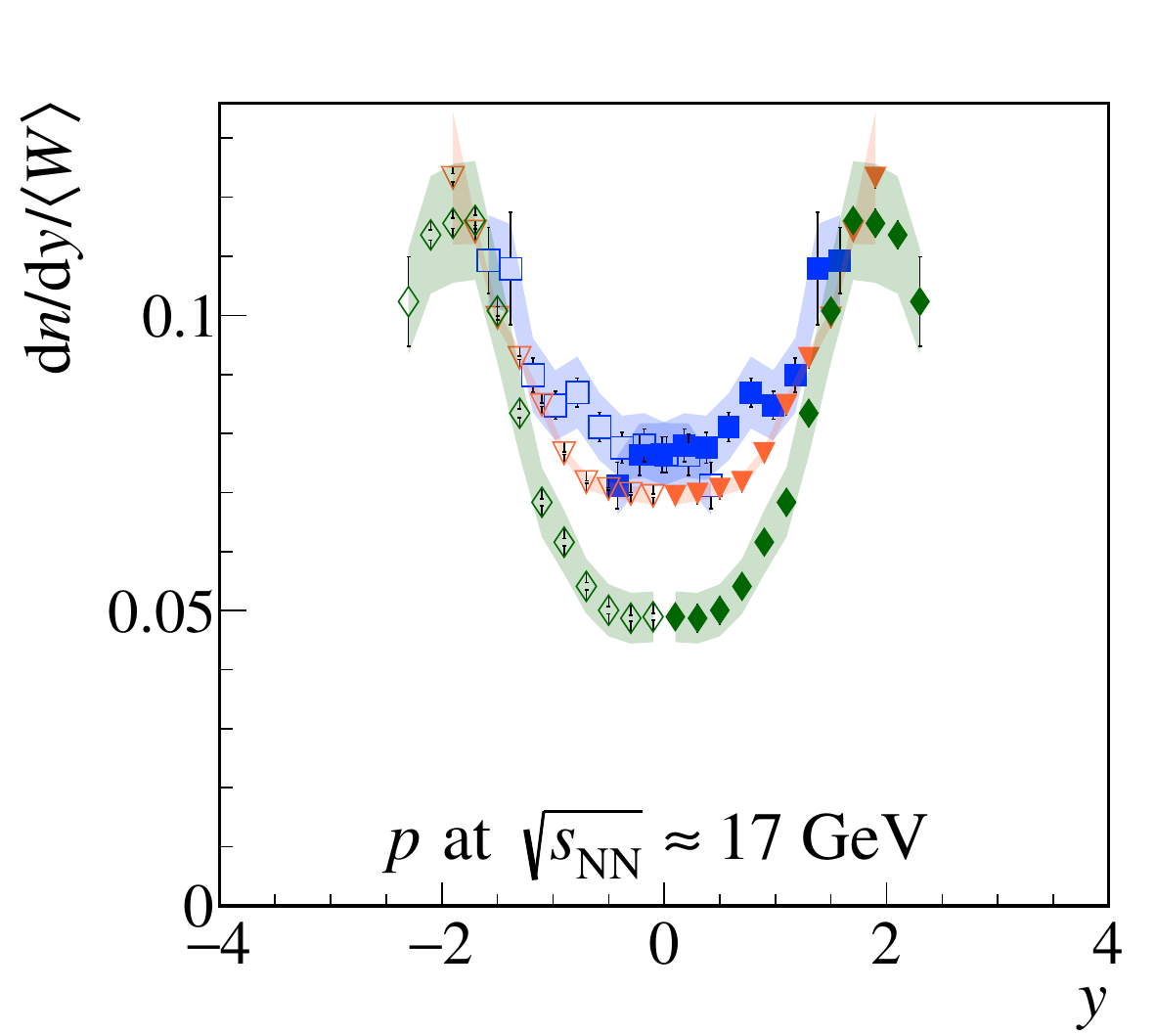}
\end{minipage}
\includegraphics[width=0.4\textwidth]{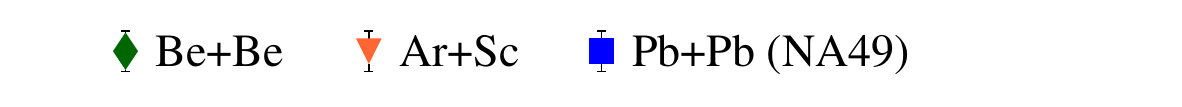}
\vspace{-0.5cm}
\caption{Comparison of normalized rapidity distributions $(\text{d}n/\text{d}y)/\langle W \rangle$ of protons produced in central Be+Be, Ar+Sc, and Pb+Pb collisions at $\sqrt{s_{\mathrm{NN}}}\approx$~5, 6, 8, 9, 12, and 17~GeV.
Statistical uncertainties are shown with error bars, and systematic uncertainties are shown as shaded bands.
Published data from Refs.~\cite{NA61SHINE:2020czq, NA61SHINE:2023epu, NA49:2011pbpb}.}
\label{fig:p_spectra}
\end{figure*}

Baryon transport and stopping are studied through the rapidity spectra of protons.
Figure~\ref{fig:p_spectra} presents the normalized rapidity distributions $(\text{d}n/\text{d}y)/\langle W \rangle$ of protons in central collisions at $\sqrt{s_{\mathrm{NN}}}\approx$~5, 6, 8, 9, 12, and 17~GeV.
With increasing collision energy, the proton yield at mid-rapidity decreases, indicating a decrease in the degree of baryon stopping and a shift of baryon number towards the fragmentation regions.
Conversely, at a fixed energy, the baryon density at mid-rapidity increases with system size.
These results suggest that baryon transport is highly sensitive to both collision energy and initial system geometry.

\vspace{-0.2cm}
\section{Lambda baryon production in central Ar+Sc collisions}
\label{sec:lambda}
New preliminary results on $\Lambda$ baryon production in the 10\% most central Ar+Sc collisions at $\sqrt{s_{\mathrm{NN}}}\approx$~5~GeV are presented along with earlier results at five higher energies $\sqrt{s_{\mathrm{NN}}}\approx$~6, 8, 9, 12, and 17~GeV previously shown in Ref.~\cite{Balkova:2025dpf}.
$\Lambda$ hyperons are reconstructed via their dominant weak decay channel $\Lambda \to p + \pi^-$ with a branching ratio of $64.1\%$.
The results discussed below refer to hadrons produced in strong interactions and electromagnetic decays.
The raw yields are corrected for detector acceptance, reconstruction efficiency, selection cuts, branching ratio, and feed-down from heavier hyperons~\cite{Balkova:2022fps}.





\begin{figure*}[ht]
\centering
\begin{minipage}{0.30\textwidth}
  \includegraphics[width=\textwidth]{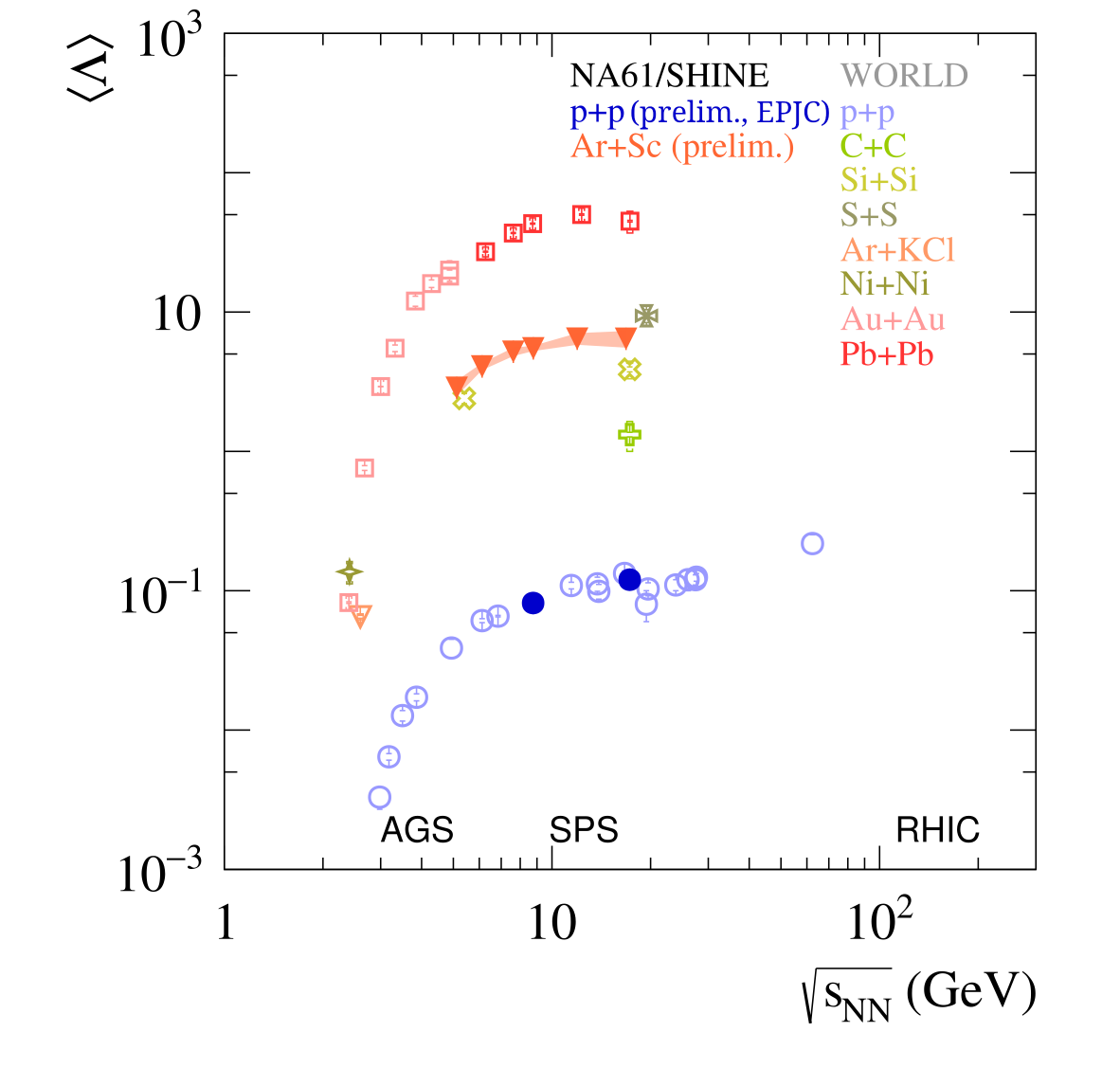}
\end{minipage}\hfill
\begin{minipage}{0.30\textwidth}
  \includegraphics[width=\textwidth]{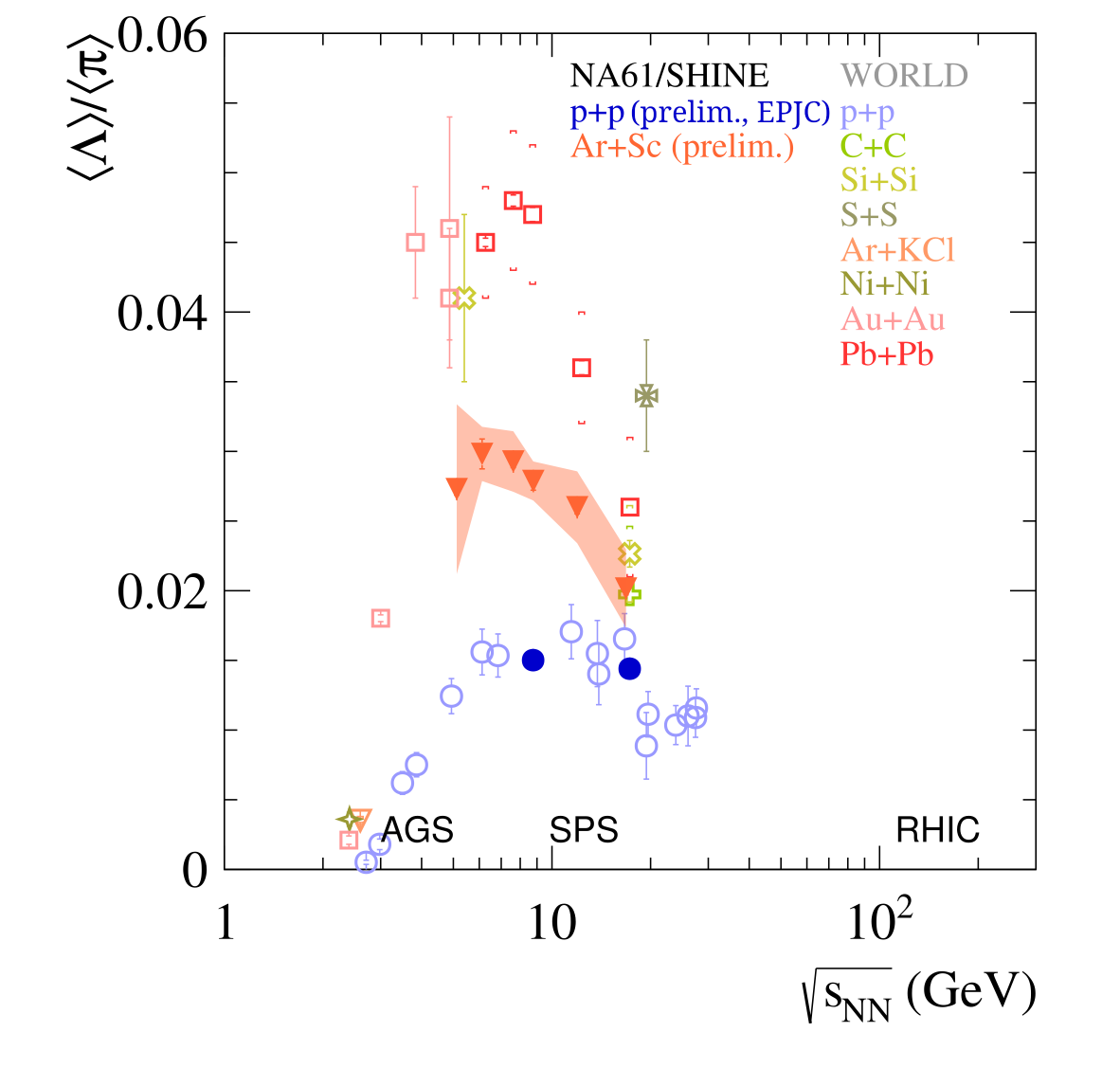}
\end{minipage}\hfill
\begin{minipage}{0.30\textwidth}
  \includegraphics[width=\textwidth]{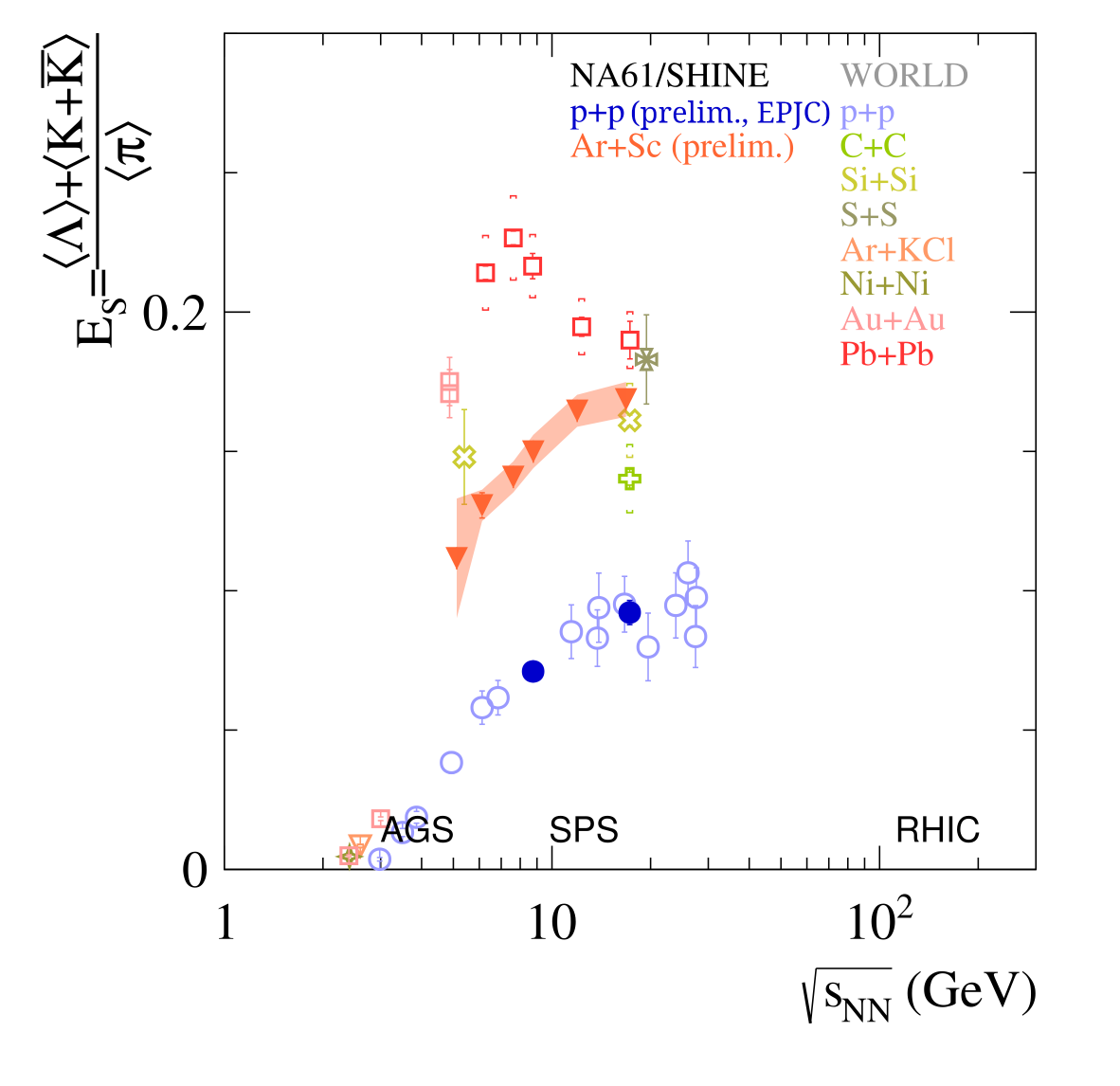}
\end{minipage}
\vspace{-0.5cm}
\caption{The energy dependence of the mean multiplicity of $\Lambda$ baryons (\textit{left}), $\langle \Lambda \rangle / \langle \pi \rangle$ ratio (\textit{middle}) and strangeness enhancement $E_S$ (\textit{right}).
The systematic uncertainties of the Ar+Sc results are presented as shaded bands.
Results for \textit{p+p}, Ar+Sc, Ar+KCl, C+C, Si+Si, S+S, Ni+Ni, Au+Au, and Pb+Pb are shown (NA61/SHINE: \textit{p+p}, Ar+Sc; NA49: C+C, Si+Si, Pb+Pb; NA35: S+S~\cite{Gazdzicki:1996pk}; NA57: Pb+Pb; STAR: Au+Au; PHENIX: Au+Au; E891: Au+Au; E895: Au+Au; E896: Au+Au; HADES: Ar+KCl, Au+Au; FOPI: Ni+Ni~\cite{FOPI:2007}; bubble chamber experiments: \textit{p+p}).
For the references to the world data, see Ref.~\cite{Balkova:2025dpf}.}
\label{fig:lambda_ratios}
\end{figure*}

The energy dependence of the mean multiplicity $\langle\Lambda\rangle$, the $\langle\Lambda\rangle/\langle\pi\rangle$ ratio, and the strangeness enhancement factor $E_S$ are shown in Figure~\ref{fig:lambda_ratios}.
The multiplicity of $\Lambda$ in Ar+Sc approaches that in Pb+Pb and is far above the \textit{p+p} baseline.
The overall trend in $\langle\Lambda\rangle/\langle\pi\rangle$ in Ar+Sc is similar to that in Pb+Pb collisions.
The $E_S$ factor is defined following Refs.~\cite{NA35:1994veo, Gazdzicki:1996pk} as $E_S = \frac{\langle\Lambda\rangle + \langle K+\overline{K}\rangle}{\langle\pi\rangle}$, where $\langle\pi\rangle = 1.5 \cdot (\langle\pi^+\rangle + \langle\pi^-\rangle)$, and the strangeness yield $\langle K+\overline{K}\rangle$ is calculated as $2 \cdot (\langle K^+\rangle + \langle K^-\rangle)$ for nucleus-nucleus collisions and as $4 \cdot \langle K^0_S \rangle$ for \textit{p+p} collisions.
It shows a pronounced peak in central Pb+Pb collisions, which is absent in Ar+Sc collisions.
This suggests that while absolute strangeness production in intermediate systems like Ar+Sc is significantly enhanced relative to the \textit{p+p} baseline, the non-monotonic energy dependence of the strangeness-to-entropy ratio remains a characteristic feature unique to the heaviest colliding systems.

\section{Isospin-symmetry breaking: excess of charged over neutral kaons}
\label{sec:kaons}
Under the assumption of exact isospin symmetry, and for collisions involving nuclei with equal numbers of protons and neutrons ($Z = N$), the production of charged kaons ($K^+, K^-$) and neutral kaons ($K^0_S$, $K^0_L$) is expected to be equal.
Specifically, for nearly isospin-symmetric nuclei such as $^{40}_{18}$Ar and $^{45}_{22}$Sc (number of $u$ and $d$ valence quarks equal within 6\%), the charged-to-neutral kaon ratio defined as $R_K = \frac{K^+ + K^-}{2 K^0_S}$ is expected to be close to unity (or slightly below due to the neutron excess, $N > Z$).

\begin{figure*}[ht]
\centering
\begin{minipage}{0.48\textwidth}
  \centering
  \includegraphics[height=0.48\textwidth]{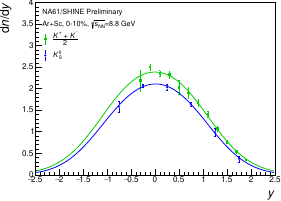}
\end{minipage}~
\begin{minipage}{0.48\textwidth}
  \centering
  \includegraphics[height=0.48\textwidth]{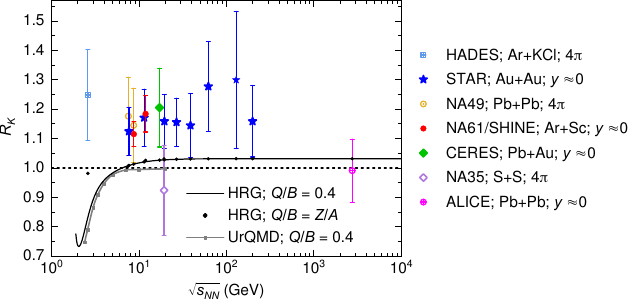}
\end{minipage}
\caption{\textit{Left}: The rapidity spectrum of $K^0_S$ mesons (blue circles, preliminary) and the averaged spectrum of charged kaons (green squares, Ref.~\cite{NA61SHINE:2023epu}) measured in 0--10\% central Ar+Sc collisions at $\sqrt{s_{\mathrm{NN}}}\approx$~9~GeV.
Total uncertainties are drawn.
Blue and green curves show the result of a double Gaussian fit to the $K^0_S$ and averaged fit for charged kaons spectra, respectively.
\textit{Right}: Charged-to-neutral kaon ratio as a function of collision energy.
Experimental data are shown by symbols with total uncertainties.
HRG baseline for electric-to-baryon charge $Q/B = 0.4$ is shown by a black line.
HRG baseline for $Q/B$ values specified according to the given types of colliding nuclei is represented by black dots.
UrQMD model results are shown by grey squares.
Figure from Ref.~\cite{NA61SHINE:2023azp} updated with the new NA61/SHINE data point.}
\label{fig:rk_ratio}
\end{figure*}

Figure~\ref{fig:rk_ratio} shows the rapidity spectrum of $K^0_S$ and charged kaons at $\sqrt{s_{\mathrm{NN}}}\approx$~9~GeV (left) and the energy/system-size dependence of the $R_K$ ratio (right).
The experimental data for central Ar+Sc collisions show a significant excess of charged kaons over neutral kaons, with $R_K \approx 1.12 - 1.18$~\cite{NA61SHINE:2023azp,Balkova:2025dpf}.

This anomalous excess represents a significant deviation from the Hadron Resonance Gas (HRG) model prediction at $75A$~GeV/$c$, which increases further when combining the $40A$ and $75A$~GeV/$c$ results.
Phenomenological models such as UrQMD~\cite{UrQMD_Ref} and HRG~\cite{HRG_Ref}, incorporating known isospin-breaking effects to different extents (e.g., quark mass differences and resonance decay branching ratios), fail to reproduce the observed large deviation of $R_K$ from unity.
As discussed in Ref.~\cite{NA61SHINE:2023azp}, at higher energies, the $R_K$ ratio in HRG is systematically higher than the one predicted by UrQMD, likely caused by UrQMD assuming $\phi$-meson decays to be exactly isospin symmetric instead of taking the branching ratios from PDG (i.e., isospin-asymmetric decays of $\phi$ are included in HRG, but not in UrQMD).
The presented experimental results provide evidence for an unexpectedly large violation of isospin symmetry in high-energy collisions of atomic nuclei~\cite{NA61SHINE:2023azp}, presenting a theoretical challenge discussed in detail in Ref.~\cite{Brylinski:2023nrb}.
Similar anomalies have also been observed in $\pi^-$+C collisions at 158 and 350~GeV/$c$~\cite{piCRef}, presenting a fundamental challenge to current particle production models and highlighting the need for a deeper understanding of soft-QCD dynamics.

\bibliographystyle{elsarticle-num}
\bibliography{sqm2026_template}

\end{document}